\overfullrule=0pt
\input harvmac

\def\a{{\alpha}}

\def\l{{\lambda}}

\def\b{{\beta}}

\def\g{{\gamma}}

\def\half{{1\over 2}}
\def\p{{\partial}}

\def\t{{\theta}}

\def\bar{\overline}
\def\zb{{\bar z}}


\Title{\vbox{\hbox{IFT-P.001/2008}
\hbox{UCLA/08/TEP/1}}}
{\vbox{
	\centerline{\bf Pure Spinor Superspace Identities for}
	\centerline{\bf Massless Four-point Kinematic Factors}
 }
}

\bigskip\centerline{Carlos R. Mafra\foot{email: crmafra@ift.unesp.br,
crmafra@physics.ucla.edu}}

\bigskip
\centerline{\it Instituto de F\'\i sica Te\'orica, State University of S\~ao Paulo}
\centerline{\it Rua Pamplona 145, 01405-900, S\~ao Paulo, SP, Brasil}
\bigskip
\centerline{\it Department of Physics and Astronomy}
\centerline{\it University of California, Los Angeles, CA 90095}

\vskip .3in Using the pure spinor formalism we prove identities which
relate the tree-level, one-loop and two-loop kinematic factors for
massless four-point amplitudes. From these identities it follows
that the complete supersymmetric one- and two-loop amplitudes
are immediately known once the tree-level kinematic factor is evaluated. In 
particular, the 
two-loop equivalence with the RNS formalism (up to an overall coefficient)
is obtained as a corollary.

\vskip .3in

\Date {January 2008}

\newsec{Introduction }

Despite numerous successes in the non-perturbative front, perturbative studies
of superstring theory remain an important 
subject of its own. For example, by computing graviton scattering amplitudes one 
can catch a glimpse of quantum gravity at work in its deepest realms, from which
one can derive superstring modifications of Einstein's equations \ref\GrossWitten{
D.~J.~Gross and E.~Witten,
  ``Superstring Modifications Of Einstein's Equations,''
  Nucl.\ Phys.\  B {\bf 277}, 1 (1986).
}.

Since the birth of the pure spinor formalism \ref\superpoincare{
	N.~Berkovits,
        ``Super-Poincare covariant quantization of the superstring,''
        JHEP {\bf 0004}, 018 (2000)
        [arXiv:hep-th/0001035].
}, the task of computing superstring scattering amplitudes has been
made easier
\ref\multiloop{
	N.~Berkovits,
        ``Multiloop amplitudes and vanishing theorems using the pure spinor
  	formalism for the superstring,''
  	JHEP {\bf 0409}, 047 (2004)
  	[arXiv:hep-th/0406055].
}\ref\topo{
	N.~Berkovits,
        ``Pure spinor formalism as an N = 2 topological string,''
  	JHEP {\bf 0510}, 089 (2005)
  	[arXiv:hep-th/0509120].
}\ref\nekrasov{
	N.~Berkovits and N.~Nekrasov,
        ``Multiloop superstring amplitudes from non-minimal pure spinor formalism,''
  	JHEP {\bf 0612}, 029 (2006)
  	[arXiv:hep-th/0609012].
}. Its main advantage over the standard Ramond-Neveu-Schwarz (RNS) and 
Green-Schwarz (GS) formulations is due to covariant quantization being possible
while having manifest space-time supersymmetry. These properties help simplify 
the required computations, which can be done efficiently.
However, until its origins are fully explained \ref\natnovo{
	N.~Berkovits,
  	``Explaining the Pure Spinor Formalism for the Superstring,''
  	[hep-th] arXiv:0712.0324.
}\ref\dafni{
	N.~Berkovits and D.~Z.~Marchioro,
        ``Relating the Green-Schwarz and pure spinor formalisms for the
  	superstring,''
  	JHEP {\bf 0501}, 018 (2005)
	[arXiv:hep-th/0412198].
}\ref\aisaka{
	Y.~Aisaka and Y.~Kazama,
  	``Origin of pure spinor superstring,''
  	JHEP {\bf 0505}, 046 (2005)
  	[arXiv:hep-th/0502208].
} and a general proof of equivalence with the RNS and GS
formalisms is obtained,
it is a good measure to check results against the established ones \ref\schwloop{
	J.~H.~Schwarz,
  	``Superstring Theory,''
  	Phys.\ Rept.\  {\bf 89}, 223 (1982).
}\ref\dhoker{
	E.~D'Hoker and D.~H.~Phong,
  	Nucl.\ Phys.\  B {\bf 715}, 3 (2005)
  	[arXiv:hep-th/0501197] 
	E.~D'Hoker and D.~H.~Phong,
  	Function,''
  	Nucl.\ Phys.\  B {\bf 715}, 91 (2005)
 	 [arXiv:hep-th/0501196].
	E.~D'Hoker and D.~H.~Phong,
  	Nucl.\ Phys.\  B {\bf 639}, 129 (2002)
  	[arXiv:hep-th/0111040].
  	E.~D'Hoker and D.~H.~Phong,
  	Nucl.\ Phys.\  B {\bf 636}, 61 (2002)
  	[arXiv:hep-th/0111016].
  	E.~D'Hoker and D.~H.~Phong,
  	Nucl.\ Phys.\  B {\bf 636}, 3 (2002)
  	[arXiv:hep-th/0110283].
  	E.~D'Hoker and D.~H.~Phong,
  	Phys.\ Lett.\  B {\bf 529}, 241 (2002)
  	[arXiv:hep-th/0110247].
}. As
far as scattering amplitudes are concerned, these checks have already successfully been done
at tree-level \ref\NatVal{
	N.~Berkovits and B.~C.~Vallilo,
  	``Consistency of super-Poincare covariant superstring tree amplitudes,''
        JHEP {\bf 0007}, 015 (2000)
        [arXiv:hep-th/0004171].
}, one-loop \ref\umloop{
	C.~R.~Mafra,
  	``Four-point one-loop amplitude computation in the pure spinor formalism,''
	JHEP {\bf 0601}, 075 (2006)
	[arXiv:hep-th/0512052].
} and
two-loops \ref\twolooptwo{
	N.~Berkovits and C.~R.~Mafra,
  	``Equivalence of two-loop superstring amplitudes in the pure spinor and  RNS
	formalisms,''
	Phys.\ Rev.\ Lett.\  {\bf 96}, 011602 (2006)
	[arXiv:hep-th/0509234].
}.

Although the scattering amplitudes were shown to be equivalent, there is a
huge difference in the amount of work required to obtain them, most notably
in the two-loop case. The RNS two-loop computations of \dhoker\ span hundreds of pages
because of complications related to the lack of manifest space-time
supersymmetry, whereas it was
obtained rather quickly in a seven-pages-long paper by Berkovits \ref\twoloop{
	N.~Berkovits,
        ``Super-Poincare covariant two-loop superstring amplitudes,''
        JHEP {\bf 0601}, 005 (2006)
        [arXiv:hep-th/0503197].
}.
Furthermore, the computation of \twoloop\ is
manifestly supersymmetric and therefore contains 
the result for all possible combination of external
states related by supersymmetry, in deep contrast with the
bosonic-only computation of \dhoker. 

One of the features of the
pure spinor formalism that makes this simplification
possible is the {\it pure spinor superspace} nature of its kinematic
factors. In this paper we will explore this property and show 
how careful manipulations
in pure spinor superspace can provide even further simplification 
of superstring scattering amplitude results. The upshot is that
the one- and two-loop kinematic factors are completely
determined (for Neveu-Schwarz and Ramond external states) 
once the tree-level amplitude is evaluated.

The massless four-point kinematic factors for the one- and two-loop 
amplitudes  are given by \multiloop\umloop\twolooptwo\
\eqn\um{
K_{\rm 1} =
\langle (\l A^1)(\l \g^m W^2)(\l \g^n W^3){\cal F}^4_{mn}\rangle
+ \rm{cycl.(234)},
}
\eqn\dois{
K_{\rm 2} =
\langle (\l \g^{mnpqr}\l){\cal F}^1_{mn}{\cal F}^2_{pq}{\cal F}^3_{rs}
(\l \g^s W^4)\rangle\Delta(1,3)\Delta(2,4)
+ \rm{perm.(1234)},
}
where $\Delta(i,j)\equiv \Delta(z_i,z_j)$ is
the basic biholomorphic 1-form defined in \dhoker.
Interestingly, the analogous expression for the tree-level amplitude
has never been found, even after the explicit computations of \ref\poli{
	G.~Policastro and D.~Tsimpis,
  	``R**4, purified,''
  	Class.\ Quant.\ Grav.\  {\bf 23}, 4753 (2006)
  	[arXiv:hep-th/0603165].
}. So in section 2 we compute
the massless four-point amplitude at tree-level to get
\eqn\zero{
K_{\rm 0}  = \half k^m_1k_2^n \langle(\l A^1)(\l A^2)(\l A^3){\cal F}^4_{mn}\rangle
-(k^1\cdot k^3)\langle A^1_n (\l A^2)(\l A^3)(\l \g^n W^4)
\rangle +(1\leftrightarrow 2).
}
Then we proceed to show in sections 3 and 4 that $K_0$, $K_1$ and $K_2$ satisfy
the following identities
\eqn\arvu{
K_{\rm 0} = -\langle (\l A^1)(\l\g^m W^2)(\l\g^n W^3){\cal F}^4_{mn}\rangle
          = -{1\over 3} K_{\rm 1}
}
\eqn\idt{
K_{\rm 2} = -32 K_0 \left[(u-t)\Delta(1,2)\Delta(3,4) 
+(s-t)\Delta(1,3)\Delta(2,4) + (s-u)\Delta(1,4)\Delta(2,3)
\right]
}

Finally, in section 5 we explain why some fermionic results reported 
in \ref\stahn{
 	C.~Stahn,
   	``Fermionic superstring loop amplitudes in the pure spinor formalism,''
   	JHEP {\bf 0705}, 034 (2007)
   	[arXiv:0704.0015 [hep-th]].
} contradict the above identities by pointing out a mistake made in \stahn\ 
which invalidates its conclusions. After clarifying this issue, we 
compute the tree-level kinematic factor \zero\ for Neveu-Schwarz and Ramond
external states, which by the identities \arvu\ and \idt\ also completely
determine the one- and two-loop kinematic factors simultaneously.

\newsec{Massless four-point amplitude at tree-level}

Although the massless four-point 
amplitude at tree-level was already explicitly computed 
in \poli, their derivation overlooked some identities in pure spinor superspace and
hid the simplicity of the result.
So in this section we compute the 
closed massless four-point amplitude at tree-level 
and extract the pure spinor superspace expression for its kinematic factor.

There are at least two motivations to pursue this goal. One is to neatly summarize
the whole tree-level computation in one pure spinor superspace expression,
which can later be used to find a relation with the one-loop amplitude,
as will be done in section 3. The other motivation is related to 
the ectoplasm method of \ref\ect{
 	N.~Berkovits,
         ``Explaining pure spinor superspace,''
         arXiv:hep-th/0612021.
}, in which pure spinor superspace expressions 
in flat space can be used to find supersymmetric invariants 
in curved space backgrounds.

Following the tree-level prescription of \superpoincare\topo, the amplitude
to compute is 
\eqn\amp{
{\cal A} = \langle V^1(z_1,{\bar z_1})V^2(z_2,{\bar z_2})V^3(z_3,{\bar z_3})
\int_{C}d^2 z_4 U(z_4,{\bar z}_4) 
\rangle.
}
The closed string vertices are given by the holomorphic square of
the open string vertices,
$V(z,{\bar z}) = {\rm e}^{ik\cdot X}\l^{\a}{\bar \l}^{\b}A_{\a}(\t)
{\bar A}_{\b}(\t)$ and $U(z,{\bar z}) = {\rm e}^{ik\cdot X}U(z){\bar U}({\bar z})$,
where \multiloop,
\eqn\integrado{
U(z) = \p\t^{\a}A_{\a}(\t) + A_m(\t)\Pi^m + d_{\a}W^{\a}(\t) 
+ \half N_{mn}{\cal F}^{mn}(\t).
}
We note that standard SL(2,C) invariance allows us to fix $z_1=0, z_2=1$ and $z_3=\infty$,
so $\langle\prod_{i=1}^4 :{\rm e}^{ik^i\cdot X(z_i,{\bar z}_i)}:\rangle
= |z_4|^{-\half \a' t}|1-z_4|^{-\half \a' u} \equiv M(z_4,{\bar z}_4)$.
The first term of \integrado\ does not contribute, while the second
gives\foot{To avoid the pollution of notation, we mostly omit the $z_i$ dependence in
the superfields even when computing OPE's, as it does not prevent the proper
understanding of the formul{\ae}.}
\eqn\triv{
\langle
A^4_m\Pi^m(z_4) \prod_{j=1}^4 :{\rm e}^{ik^j\cdot X(z_j,{\bar z}_j)}:
\rangle = \sum_{j=1}^3 {\a'\over 2}{ik^m_j \over z_j-z_4}
\langle 
(\l A^1)(\l A^2)(\l A^3)A^4_m
\rangle M(z_4,{\bar z}_4).
}
Using the standard OPE's \multiloop\
\eqn\opes{
N^{mn}(z_4)\l^{\a}(z_j) =
{\a'\over 4}{(\l\g^{mn})^{\a} \over z_j-z_4}, \quad
d_{\a}(z_4)V(z_j) = -{\a'\over 2}{D_{\a}V \over z_j - z_4},
}
we obtain the following OPE identity:
$$
\langle
(\l A^1)(\l A^2)(\l A^3)\left(
d_{\a}(z_4)W_4^{\a} +\half N^{mn}(z_4){\cal F}_{mn}^4 \right)
\rangle = 
$$
\eqn\pilita{
= {\a' \over 2(z_1-z_4)}\langle
A^1_m (\l A^2)(\l A^3)(\l \g^m W^4) 
\rangle - (1\leftrightarrow 2) + (1\leftrightarrow 3).
}
To show this, one uses \opes\ to get
$$
\langle
(\l A^1)(z_1)(\l A^2)(z_2)(\l A^3)(z_3)d_{\a}(z_4)W_4^{\a}
\rangle = 
$$
$$
{\a'\over 2(z_1-z_4)}\langle
D_{\a}(\l A^1)(\l A^2)(\l A^3) W_4^{\a}
\rangle - (1\leftrightarrow 2) + (1\leftrightarrow 3).
$$
Concentrating for simplicity on the first term, the use of the
super-Yang-Mills identity 
$D_{\a}(\l A) = -(\l D)A_{\a} + (\l\g^m)_{\a}A_m$
allows the numerator to be rewritten as
\eqn\tmp{
\langle
D_{\a}(\l A^1)(\l A^2)(\l A^3) W_4^{\a}
\rangle = - \langle
(\l D A^1_{\a})(\l A^2)(\l A^3) W_4^{\a}
\rangle + \langle
A^1_m(\l A^2)(\l A^3) (\l\g^m W^4)
\rangle.
}
As BRST-exact terms decouple, the first term in the right hand side of \tmp\
becomes
$$
-{\a'\over 2(z_1-z_4)} \langle
(\l D A^1_{\a})(\l A^2)(\l A^3) W_4^{\a}
\rangle = 
-{\a'\over 2(z_1-z_4)} \langle
A^1_{\a}(\l A^2)(\l A^3) (\l D)W_4^{\a}
\rangle 
$$
$$
= -{\a'\over 8(z_1-z_4)}\langle
(\l \g^{mn} A^1)(\l A^2)(\l A^3){\cal F}_{mn}^4
\rangle.
$$
However, this term is exactly canceled by the $(z_1-z_4)^{-1}$ contribution from
the OPE
$$
\half\langle
(\l A^1)(\l A^2)(\l A^3)
(N^{mn}{\cal F}^4_{mn})
\rangle = {\a'\over 8(z_1-z_4)} \langle
(\l \g^{mn} A^1)(\l A^2)(\l A^3){\cal F}_{mn}^4
\rangle +{\ldots} ,
$$
which finishes the proof of \pilita.

With the results \triv\ and \pilita, the correlation in 
the amplitude \amp\ reduces to 
$$
{\cal A} = \left({\a'\over 2} \right)^2 \int_{C}d^2z_4
\left(
{F_{12}\over z_4} + {F_{21}\over 1-z_4}
\right)\left(
{{\bar F}_{12}\over\zb_4} + {{\bar F}_{21}\over 1-\zb_4}
\right)|z_4|^{-\half \a' t}|1-z_4|^{-\half \a' u},
$$
where $F_{12} = ik_1^m\langle (\l A^1)(\l A^2)(\l A^3) A^4_m
\rangle + \langle A^1_m (\l A^2)(\l A^3)(\l\g^m W^4)
\rangle$ and $F_{21}$ is obtained by exchanging $1\leftrightarrow 2$.
The integral can be evaluated using the following formula \ref\sduality{
	E.~D'Hoker, M.~Gutperle and D.~H.~Phong,
  	``Two-loop superstrings and S-duality,''
  	Nucl.\ Phys.\  B {\bf 722}, 81 (2005)
  	[arXiv:hep-th/0503180].
}
$$
\int_{C} d^2z z^N(1-z)^M \zb^{\bar N}(1-\zb)^{\bar M} =
2\pi {\Gamma(1+N)\Gamma(1+M)  \over \Gamma(2+N+M) }
{\Gamma(-1-{\bar N}-{\bar M})  \over \Gamma(-{\bar N})\Gamma(-{\bar M}) }.
$$
After a few manipulations one finally gets
$$
{\cal A} = -2\pi ({\a'\over 2})^4 K_0{\bar K}_0
{\Gamma(-{\displaystyle \a' t\over 4})\Gamma(-{\displaystyle \a' u\over 4})
\Gamma(-{\displaystyle \a' s\over 4}) 
\over \Gamma(1+{\displaystyle \a' t\over 4})\Gamma(1+{\displaystyle \a' u\over 4})
\Gamma(1+{\displaystyle \a' s\over 4})},
$$
where $K_0 = \half(u F_{12} + t F_{21})$ is given by
\eqn\kin{
K_0  = 
\langle  (\p_m A^1_n) (\l A^2)\p^m(\l A^3)(\l \g^n W^4)\rangle
- \half\langle \p^m(\l A^1)\p^n(\l A^2)(\l A^3){\cal F}^4_{mn} \rangle
+ (1\leftrightarrow 2),
}
which is the sought-for kinematic factor in pure spinor superspace.
As will become clear later, it is convenient to rewrite \kin\ without
explicit labels,
\eqn\this{
K_0 = 2\langle (\p_m A_n) (\l A)\p^m(\l A)(\l \g^n W) \rangle
- \langle (\l A)\p^m(\l A)\p^n(\l A){\cal F}_{mn}\rangle. 
}
Furthermore, using the identities of \ref\NMPS{
 	N.~Berkovits and C.~R.~Mafra,
         ``Some superstring amplitude computations with the non-minimal pure spinor
         formalism,''
   	JHEP {\bf 0611}, 079 (2006)
   	[arXiv:hep-th/0607187].
} we will compute in section 5 the whole component expression of 
\kin\ (for Neveu-Schwarz and Ramond external states).
We will see that expression \kin\ neatly summarizes the rather lenghty
computations of \poli.

\newsec{Relating tree-level and one-loop kinematic factors}

Using the well-known superfield equations of motion in the 
formulation of ten-dimensional 
Super-Yang-Mills theory in superspace \ref\witten{
	E.~Witten,
        ``Twistor - Like Transform In Ten-Dimensions,''
        Nucl.\ Phys.\ B {\bf 266}, 245 (1986).
}\ref\ictp{
	N.~Berkovits,
        ``ICTP lectures on covariant quantization of the superstring,''
	arXiv:hep-th/0209059.
}, one can show that
\eqn\SYM{
Q{\cal F}_{mn} = 2\p_{[m} (\l\g_{n]} W), \quad
Q W^{\a} = {1\over 4}(\l\g^{mn})^{\a}{\cal F}_{mn},\quad  
QA_m = (\l\g_m W) + \p_m(\l A),
}
where $Q=\oint \l^{\a}d_{\a}$ is the pure spinor BRST operator\foot{We
refer the reader to the lectures notes in the pure
spinor formalism \ictp\ for the basic definitions.}. With these
relations in hand we will show that \arvu\ holds true.
To prove this we note that $\langle (\l A)\p^m(\l A)(QA^n)F_{mn}
\rangle = -\langle (\l A)\p^m(\l A)A^n (QF_{mn})\rangle $, which upon use of
\SYM\ and momentum conservation becomes
$$
\langle (\l A)\p^m(\l A)(QA^n)F_{mn}
\rangle =  \langle  (\l A)\p^m(\l A) \p_m A_n  (\l \g^n W) \rangle
$$
\eqn\part{
- \langle  \p_n (\l A)\p_m(\l A) A^n  (\l \g^m W) \rangle
- \langle   (\l A)\p_n\p_m(\l A) A^n  (\l \g^m W) \rangle.
}
The second term can be rewritten like
$$
\langle  \p_n (\l A)\p_m(\l A) A^n  (\l \g^m W) \rangle =
- \langle  (\l A)(\l \g^m W)\left[ 
A^n \p_m\p_n(\l A) + \p^n(\l A)\p_m A_n 
\right]\rangle
$$
as can be shown by integrating $\p^m$ by parts and 
using the equation of motion for $W^{\a}$. So,
$$
\langle (\l A)\p^m(\l A)(QA^n)F_{mn} \rangle =
\langle (\l A)\p^m(\l A)(\l\g^n W)F_{mn} \rangle 
- 2 \langle   (\l A)\p_n\p_m(\l A) A^n  (\l \g^m W) \rangle
$$
which implies that
$
\langle (\l A)\p^m(\l A)\p^n(\l A)F_{mn} \rangle =
-2 \langle   (\l A)\p_n\p_m(\l A) A^n  (\l \g^m W) \rangle$, or equivalently,
\eqn\impli{
\langle (\l A)\p^m(\l A)\p^n(\l A)F_{mn} \rangle = 
-2 \langle   (\l A)\p_n(QA_m) A^n  (\l \g^m W) \rangle.
}
Using $[Q,\p^n]=0$ and the decoupling of BRST-trivial operators, equation \impli\
becomes
$$
\langle (\l A)\p^m(\l A)\p^n(\l A)F_{mn} \rangle =
2\langle (\l A)(\p_n A_m) (QA^n)  (\l \g^m W) \rangle
$$
\eqn\segue{
= \langle (\l A)(\l\g^m W)(\l \g^n W)F_{mn} \rangle +
2\langle (\p_n A_m)(\l A)\p^n(\l A) (\l\g^m W) \rangle.
}
Plugging \segue\ in the tree-level kinematic factor \this\
we finally obtain
\eqn\mapocha{
K_0 = - \langle (\l A)(\l\g^m W)(\l \g^n W)F_{mn} \rangle = -{1\over 3}K_1,
}
which finishes\foot{
This proof was completed a few days after being told that
Paul Howe had independently shown the same thing \ref\howe{Paul Howe, private 
communication.}.
} the proof of \arvu.

\newsec{Relating one- and two-loop kinematic factors}

To obtain a relation between the one- and two-loop kinematic
factors we first need to show that
$\langle (\l A^1)(\l \g^m W^2)(\l \g^n W^3){\cal F}^4_{mn}
\rangle$ is completely symmetric in the labels $(1234)$. This can be
done by noting that\foot{I thank Nathan Berkovits for suggesting (4.1) to me.},
\eqn\brst{
\langle (\l \g^{mnpqr} \l)(\l A^1)(W^2 \g_{pqr} W^3){\cal F}^4_{mn}
\rangle =
4 \langle (\l A^1)Q\left[(W^2 \g_{pqr} W^3)\right](\l \g^{pqr} W^4)
\rangle.
}
Together with the identities $(\l\g^{mn}\g^{pqr}W^2)(\l\g_{pqr}W^4) 
= -48 (\l\g^{[m}W^2)(\l\g^{n]}W^4)$ and 
$(\l \g^{mnpqr} \l)(W^2 \g_{pqr} W^3) = 
-96 (\l \g^{[m} W^2)(\l \g^{n]} W^3)$,  
equation \brst\ implies that
$$
\langle (\l A^1)(\l \g^m W^4)(\l \g^n W^2){\cal F}^3_{mn}\rangle +
\langle (\l A^1)(\l \g^m W^3)(\l \g^n W^4){\cal F}^2_{mn}\rangle =
$$
\eqn\cont{
= 2 \langle  (\l A^1)(\l \g^m W^2)(\l \g^n W^3){\cal F}^4_{mn}
\rangle.
}
From \cont\ it follows that,
\eqn\dem{
K_{\rm 1-loop} = 3 \langle (\l A^1)(\l \g^m W^2)(\l \g^n W^3){\cal F}^4_{mn}\rangle.
}
Furthermore, the independence of which vertex operator
we choose to be non-integrated \NatVal\  implies total
symmetry
of $\langle (\l A^1)(\l \g^m W^2)(\l \g^n W^3){\cal F}^4_{mn}
\rangle$ in the labels $(1234)$.

Now we can relate the one- and two-loop kinematic factors by noting that
$$
(\l \g^{mnpqr}\l){\cal F}^1_{mn}{\cal F}^2_{pq}{\cal F}^3_{rs}
(\l \g^s W^4) =
-4 Q\left[
(\l \g^r\g^{mn} W^2)(\l \g^s W^4){\cal F}^1_{mn}{\cal F}^3_{rs}
\right] 
$$
\eqn\note{
- 8 ik^1_m (\l\g_n W^1)(\l\g^r\g^{mn}W^2)(\l\g^s W^4){\cal F}^3_{rs},
}
where the pure spinor constraint $(\l\g^m\l)=0$ and the
identity $\eta_{mn}\g^m_{\a(\b}\g^n_{\g\delta)}=0$ must be used to show the
vanishing of terms containing factors of $(\l\g^m)_{\a}(\l\g_m)_{\b}$. 
Furthermore,
as BRST-exact terms  decouple from 
pure spinor correlations $\langle{\ldots}\rangle$,
equation \note\ implies
\eqn\inte{
\langle
(\l \g^{mnpqr}\l){\cal F}^1_{mn}{\cal F}^2_{pq}{\cal F}^3_{rs}
(\l \g^s W^4)\rangle =
+16 ik^1_m \langle 
(\l\g^r W^1)(\l\g^m W^2)(\l\g^s W^4){\cal F}^3_{rs},
\rangle,
}
where we have used $k^1_m(\l\g_n W^1)(\l\g^r\g^{mn}W^2) =
-2 k^1_m(\l\g^r W^1)(\l\g^m W^2)$, which is valid when
the equation of motion $k^1_m(\g^m W^1)_{\a}=0$ is satisfied.

Using $(\l\g_m W^2) = QA^2_m - ik^2_m(\l A^2)$ and  
$\langle (\l\g^r W^1)Q(A_2^m)(\l\g^s W^4)
{\cal F}^3_{rs}\rangle = 0$ we arrive at the following pure
spinor superspace identity
\eqn\mardelpi{
\langle
(\l \g^{mnpqr}\l){\cal F}^1_{mn}{\cal F}^2_{pq}{\cal F}^3_{rs}
(\l \g^s W^4)\rangle =
- 16 (k^1\cdot k^2) \langle 
(\l A^2)(\l\g^r W^1)(\l\g^s W^4){\cal F}^3_{rs}
\rangle
}
Multiplying \mardelpi\ by $\Delta(1,3)\Delta(2,4)$ and
summing over permutations leads to the following identity,
\eqn\nt{
K_2 = {32 \over 3}K_1 \left[(u-t)\Delta(1,2)\Delta(3,4) 
+(s-t)\Delta(1,3)\Delta(2,4) + (s-u)\Delta(1,4)\Delta(2,3)
\right],
}
where we used \dem\ and the standard Mandelstam variables
$s=-2(k^1\cdot k^2)$, $t=-2(k^1\cdot k^4)$,  $u=-2(k^2\cdot k^4)$.

In view of the results in section 5, \nt\ not only provides 
a simple proof of two-loop 
equivalence with the (bosonic) RNS result of \dhoker\ but it also
automatically implies the knowledge of the full amplitude, including
fermionic external states.

\newsec{The complete tree-level, one- and two-loop kinematic factors}

The fermionic results reported in the first version of \stahn\ are
in direct contradiction with the identities \dem\ and \nt. 
The two-loop kinematic factor (for 2F2B) was incorrectly argued to 
not have the simple form of \nt\ and the 2F2B one-loop 
computation of \stahn\ does not obey identity \dem. 
We clarify these issues by pointing out
the mistake made in \stahn\ which, strictly speaking,
invalidates all its fermionic computations. 
After these issues are settled in the next paragraph, we
compute the whole component expansion of \kin, as that
will automatically imply the full knowledge of $K_2$ for the
first time.
This is a remarkable example of
the simplifying power of the pure spinor formalism.

In \stahn\ the first component of $W^{\a}(\t)$, denoted by $u^{\a}$, is considered 
to be bosonic instead of fermionic. So, in a strict sense, all fermionic computations
in \stahn\ are unreliable and all discussions based on symmetry properties of fermionic 
kinematic factors need review. In particular, the discussion of the 2F2B kinematic
factor at two-loops is wrong because we have proven in \nt\ that it is in fact
proportional to the one-loop result. 

This symmetry mistake in \stahn\ is also apparent in its computation 
of $K_1^{\rm 2F2B}$ at one-loop. One can check it 
in the first formula of section 3.3, where the factor $(1-\pi_{34})$
should be $(1+\pi_{34})$. To see this note that the one-loop kinematic factor, with the
cyclic permutations written out explicitly,
$$
K_1 = \langle (\l A^1)(\l \g^m W^2)(\l \g^n W^3){\cal F}_{mn}^4\rangle +
\langle (\l A^1)(\l \g^m W^4)(\l \g^n W^2){\cal F}_{mn}^3\rangle +
$$
\eqn\cert{
+ \langle (\l A^1)(\l \g^m W^3)(\l \g^n W^4){\cal F}_{mn}^2\rangle,
}
can be rewritten as
$$
K_1 = \langle (\l A^1)(\l \g^m W^2)(\l \g^n W^3){\cal F}_{mn}^4\rangle +
\langle (\l A^1)(\l \g^m W^2)(\l \g^n W^4){\cal F}_{mn}^3\rangle +
$$
\eqn\certo{
+ \langle (\l A^1)(\l \g^m W^3)(\l \g^n W^4){\cal F}_{mn}^2\rangle,
}
because $(\l \g^m W^4)(\l \g^n W^2) = - (\l \g^n W^2)(\l \g^m W^4)$ due to
the fermionic nature of $W^{\a}(\t)$. So the first line of equation \certo\ can be written
in terms of the permutation symbol $\pi_{ij}$ as 
$(1+\pi_{34})\langle (\l A^1)(\l \g^m W^2)(\l \g^n W^3){\cal F}_{mn}^4\rangle$, and
not with $(1-\pi_{34})$ like shown\foot{We also note that it is not necessary 
to make distinctions between $W^{\rm (even)}$ and $W^{\rm (odd)}$ to obtain \certo. One
can choose which superfields contribute with fermions ($\chi^{\a}$)  or bosons ($e_m$) 
after these performing
these pure spinor superspace manipulations.} in \stahn.
Because of this mistake, the conclusion
reached in \stahn\ was that the first line of \cert\ vanished instead of being 
$+2\langle (\l A^1)(\l \g^m W^3)(\l \g^n W^4){\cal F}_{mn}^2\rangle$, as one would conclude
by using the identities proven in section 4. In fact, we know that all three terms in \cert\
are equal because of the total symmetry property demonstrated in this paper. So if one
subtracts two of them the answer must be zero. The fact that in \stahn\ the author
concludes 
that $(1-\pi_{34})\langle (\l A^1)(\l \g^m W^2)(\l \g^n W^3){\cal F}_{mn}^4\rangle =0$
indicates that its computer codes are indeed correct.

In the sequence we use the following ${\cal N}=1$ super-Yang-Mills $\t$ expansions
$$
A_{\a}(x,\t)={1\over 2}a_m(\g^m\t)_\a -{1\over 3}(\xi\g_m\t)(\g^m\t)_\a
-{1\over 32}F_{mn}(\g_p\t)_\a (\t\g^{mnp}\t) 
$$
$$
+ {1\over 60}(\g_m\t)_{\a}(\t\g^{mnp}\t)(\p_n\xi\g_p\t) + \ldots
$$
$$
A_{m}(x,\t) = a_m - (\xi\g_m\t) - {1\over 8}(\t\g_m\g^{pq}\t)F_{pq}
         + {1\over 12}(\t\g_m\g^{pq}\t)(\p_p\xi\g_q\t) + \ldots
$$
$$
W^{\a}(x,\t) = \xi^{\a} - {1\over 4}(\g^{mn}\t)^{\a} F_{mn}
           + {1\over 4}(\g^{mn}\t)^{\a}(\p_m\xi\g_n\t)
	   + {1\over 48}(\g^{mn}\t)^{\a}(\t\g_n\g^{pq}\t)\p_m F_{pq} 
	   + \ldots
$$
$$
{\cal F}_{mn}(x,\t) = F_{mn} - 2(\p_{[m}\xi\g_{n]}\t) + {1\over
4}(\t\g_{[m}\g^{pq}\t)\p_{n]}F_{pq} + {\ldots},
$$
and the pure spinor superspace identities in the appendix of \NMPS. Here $\xi^{\a}(x) =
\chi^{\a}{\rm e}^{ik\cdot x}$ and $a_m(x) = e_m {\rm e}^{ik\cdot x}$ describe the gluino and
gluon respectively, while $F_{mn} = 2\p_{[m} a_{n]}$ is the gluon field-strength.

Now we compute the whole component expansion of
\eqn\kint{
K_0 = \half k_1^mk_2^n \langle (\l A^1)(\l A^2)(\l A^3){\cal F}^4_{mn}\rangle
- (k^1\cdot k^3)\langle A^1_m (\l A^2)(\l A^3)(\l\g^m W^4)\rangle + (1\leftrightarrow 2).
}
The first term doesn't contribute in the computation of $K_0(f_1f_2f_3f_4)\equiv K_0^{\rm 4F}$,
while the second leads to\foot{I acknowledge the use of the GAMMA package \ref\ulf{
	U.~Gran,
  	``GAMMA: A Mathematica package for performing Gamma-matrix algebra and  Fierz
  	transformations in arbitrary dimensions,''
  	arXiv:hep-th/0105086.
} in these computations.}
$$
K_0^{\rm 4F} = -{1\over 9}(k^1\cdot k^3)\langle (\l\g^a \t)(\l\g^b \t)(\l\g^c \chi^4)
(\chi^3\g^b\t)(\t\g^c\chi^1)(\chi^2\g^a \t)\rangle +(1\leftrightarrow 2),
$$
$$
 = {1\over 5760}\left[(\chi^1\g^m\chi^2)(\chi^3\g_m\chi^4)\left[
(k^2\cdot k^3)-(k^1\cdot k^3)
\right]
-{1\over 12}(k^3\cdot k^4)(\chi^1\g^{mnp}\chi^2)(\chi^3\g_{mnp}\chi^4)
\right].
$$
Using the following Fierz identity \ref\krotov{
  	V.~Alexandrov, D.~Krotov, A.~Losev and V.~Lysov,
  	``On Pure Spinor Superfield Formalism,''
  	JHEP {\bf 0710}, 074 (2007)
  	[arXiv:0705.2191 [hep-th]].
}
$$
(\chi^1\g^{mnp}\chi^2)(\chi^3\g_{mnp}\chi^4)= 24(\chi^1\g^m\chi^3)(\chi^2\g_m\chi^4)
-12(\chi^1\g^m\chi^2)(\chi^3\g_m\chi^4),
$$ 
we arrive at
\eqn\qf{
K_0^{\rm 4F} = -{1\over 2880}\left[
(k^1\cdot k^3)(\chi^1\g^m\chi^2)(\chi^3\g_m\chi^4)+
(k^3\cdot k^4)(\chi^1\g^m\chi^3)(\chi^2\g_m\chi^4)
\right].
}
Both terms of \kint\ contribute in the $K_0^{\rm 2B2F} \equiv K_0(f_1f_2b_3b_4)$
kinematic factor,
$$
K_0^{\rm 2B2F} =
-{1\over 36} k^m_1k^n_2 F^4_{mn} e^3_p \langle
(\l\g^t \t)(\l\g^u\t)(\l\g^p\t)(\t\g_t\chi^1)(\chi^2\g_u\t)
\rangle
$$
$$
-{1\over 24}(k^1\cdot k^3)F^4_{mn} e^3_p \langle
(\l\g^t \t)(\l\g^p\t)(\l\g^q\g^{mn}\t)(\t\g_q\chi^1)(\chi^2\g_t\t)
\rangle + (1\leftrightarrow 2)
$$
\eqn\lado{
= {1\over 5760}F^4_{mn}e^3_p\left[
k^m_1k_2^n(\chi^1\g^p\chi^2) + \half(k^1\cdot k^3) (\chi^1\g^{mn}\g^{p}\chi^2)
\right] + (1\leftrightarrow 2)
}
It is worth noticing that the explicit computation of $K_0^{\rm 2B2F}$ becomes 
easier if we use the identity \mapocha\ with a convenient choice for the labels
in the right hand side, namely
$K_0 = -\langle (\l A^1)(\l\g^m W^3)(\l\g^n W^4){\cal F}^2_{mn}\rangle$, because
now one can check that only one term contributes
$$
K_0^{\rm 2B2F} = {1\over 24}\langle
(\l\g^p \t)(\l\g^{[m|}\gamma^{rs}\t)(\l\g^{|n]}\gamma^{tu}\t)(\t\g_p\chi^1)(\chi^2\g_n\t)
\rangle k^2_m F^3_{rs}F^4_{tu}
$$
\eqn\outro{
=  {1\over 5760}F^3_{mn}F^4_{rs}\left[
- i(\chi^1\g^r \chi^2)\eta^{sm}  k_2^n 
+ {i\over 2} (\chi^1\g^{mnr} \chi^2)k_2^s
\right] 
+ (3\leftrightarrow 4).
}
One can verify that \lado\ and \outro\ are in fact equal and equivalent to the RNS
result (see for example \ref\siegel{
	K.~Lee and W.~Siegel,
  	``Simpler superstring scattering,''
  	JHEP {\bf 0606}, 046 (2006)
  	[arXiv:hep-th/0603218].
}). This equality can also be regarded as a check of identity \mapocha, which
is reassuring.
The computation of $K_0^{\rm 4B}$ is straightforward (and can also be deduced
from the one-loop result of \umloop). One can in fact check that 
$$
K_0^{\rm 4B} = {1\over 5760}\left[
- \half (e^1\cdot e^3)(e^2\cdot e^4)ts
- \half (e^1\cdot e^4)(e^2\cdot e^3)us
- \half (e^1\cdot e^2)(e^3\cdot e^4)tu \right.
$$
$$
+(k^4\cdot e^1)(k^2\cdot e^3)(e^2\cdot e^4)s+
(k^3\cdot e^2)(k^1\cdot e^4)(e^1\cdot e^3)s
$$
$$
+(k^3\cdot e^1)(k^2\cdot e^4)(e^2\cdot e^3)s+
(k^4\cdot e^2)(k^1\cdot e^3)(e^1\cdot e^4)s
$$
$$
+(k^1\cdot e^2)(k^3\cdot e^4)(e^1\cdot e^3)t+
(k^4\cdot e^3)(k^2\cdot e^1)(e^2\cdot e^4)t
$$
$$
+(k^4\cdot e^2)(k^3\cdot e^1)(e^3\cdot e^4)t+
(k^1\cdot e^3)(k^2\cdot e^4)(e^1\cdot e^2)t
$$
$$
+(k^2\cdot e^1)(k^3\cdot e^4)(e^2\cdot e^3)u+
(k^4\cdot e^3)(k^1\cdot e^2)(e^1\cdot e^4)u
$$
$$
\left.
+ (k^4\cdot e^1)(k^3\cdot e^2)(e^3\cdot e^4)u+
(k^2\cdot e^3)(k^1\cdot e^4)(e^1\cdot e^2)u {\phantom\half}\right]
$$
\eqn\final{
 = {1\over 2880} t_8^{m_1n_1m_2n_2m_3n_3m_4n_4}
 F^1_{m_1n_1}F^2_{m_2n_2}F^3_{m_3n_3}F^4_{m_4n_4},
}
where we used the $t_8$ tensor definition of \ref\gswu{
	M.~B.~Green, J.~H.~Schwarz and E.~Witten,
  	``Superstring Theory. Vol. 1: Introduction,''
{\it  Cambridge, Uk: Univ. Pr. ( 1987) 469 P. ( Cambridge Monographs On Mathematical Physics)}
}\ref\gswd{
	M.~B.~Green, J.~H.~Schwarz and E.~Witten,
        ``Superstring Theory. Vol. 2: Loop Amplitudes, Anomalies And Phenomenology,''
{\it  Cambridge, Uk: Univ. Pr. ( 1987) 596 P. ( Cambridge Monographs On Mathematical Physics)}
}.

\vskip 15pt
{\bf Acknowledgements:} 
I would like to thank UCLA and especially Eric D'Hoker for the hospitality 
I enjoyed during the completion of this work. I thank Nathan Berkovits for
his faster-than-light email help when away and for his office door being
always open when at the institute. I thank Christian Stahn for
discussions regarding \stahn. I also acknowledge
FAPESP grant 04/13290-8 for financial support.

\listrefs

\end